\documentstyle[12pt]{article}
\begin{document}

{\hfill  \bf Preprint Alberta-Thy-38-94.}

\begin{center}
{ \Large \bf
On the Off-Mass-Shell Deformation \\
of the Nucleon
Structure Function.
}\\[10mm]
\noindent
  A.Yu. Umnikov and F.C. Khanna\\[4mm]
{\em
 Theoretical Physics Institute, Physics Department,
 University of
 Alberta, \\Edmonton,
  Alberta T6G 2J1,
 \\ and
 TRIUMF, 4004 Wesbrook Mall, Vancouver, B.C. V6T 2A3, Canada}\\[4mm]
 L.P. Kaptari\\[4mm]
 {\em
  Laboratory of Theoretical Physics,
Joint Institute for Nuclear Research,\\ 141980, Dubna, Russia}

 \end{center}

\begin{abstract}
 The off-mass-shell
 behavior of the nucleon structure function, $F_2^N$, is
 studied within an approach motivated by the
 Sullivan model. Deep inelastic scattering on the
 nucleon is considered in the second order in the
 pion-nucleon coupling constant,
 corresponding to the dressing of the bare nucleons
 by the mesonic cloud.
 The inclusive and semi-inclusive
 deep inelastic
 processes on the deuteron
 involving  off-shell nucleons are considered.
 A deformation of
 the mesonic cloud for the off-mass-shell
 nucleon, compared to the free one,
 generates observable effects
 in deep inelastic scattering.
 In particular,
 it leads to the breakdown of
 the convolution model,
 i.e. the deuteron structure functions
 are not expressed through the free nucleon
 structure function.
 Analysis of the semi-inclusive deep inelastic scattering on the
deuteron,
      in the spectator approximation, shows
      that this
      reaction opens  new possibilities to study the role of
      the off-shell effects in determining in detail
      the nucleon structure function.
\end{abstract}

\newpage

\section {Introduction}

 Studying of the deep inelastic scattering (DIS) of  leptons by
nuclei
 has created
 a special problem, description of the DIS on the
 off-mass-shell nucleons.
 By itself this problem is not something
 unique or new in nuclear physics, since the same problem
 of the off-mass-shell amplitudes
 arises in investigations of the reactions of any kind
 (see e.g.~\cite{defor}).

 Historically, all off-mass-shell effects in the  DIS on the
 nucleons are divided into two classes.
 First, these are the so-called "binding effects"~\cite{vag,vag1}. It
was
 noticed that compared to the free nucleon structure function
$F_2^N(x_N,Q^2)$,
 the structure function of the off-mass-shell nucleon, the "bound
nucleon", has
 a shift in the kinematical variables:
\begin{eqnarray}
 x = \frac{Q^2}{2pq} = \frac{x_N m}{p_0+p_z},
\end{eqnarray}
 where $Q^2 = -q^2$, $q$ is the 4-momentum transfer in the reaction,
 m is nucleon mass, $p$ is the virtual nucleon momentum
 and $x_N = -Q^2/(2q_0m)$ is the
 Bjorken scaling variable for free nucleon at rest. Since $p_0 < m$
 and averaging on the nucleus results in
 $\langle p_0 + p_z\rangle < m$, the
 structure function of the bound nucleon is "shifted" to a smaller
 value of $x_N$. The exception is the region of high $x_N$,
$x_N{\
\lower-1.0pt\vbox{\hbox{\rlap{$>$}\lower6pt\vbox{\hbox{$\sim$}}}}\
}1$,
 where due to the admixture of the $p_0 + p_z > m$, the "Fermi
motion",
 the bound nucleon structure function is extended beyond the single
 nucleon kinematics.
 When only such kinematical effects are taken into account
 the bound nucleon contribution to the nuclear structure function
 is given as a convolution of the free nucleon structure function
 with an effective distribution function for the
 nucleons~\cite{jaffe,muld}. Taking account of the binding effects
 in this distribution function
 leads to a description  of
 the EMC-effect in the DIS~\cite{vag,vag1}.

 Second, the off-mass-shell effects are all
 possible phenomena, other than the binding effects,
 which make the off-mass-shell
 structure function different from the on-mass-shell one.
 Such effects are related to the fact that the nucleon has an
internal
structure
 and the structure of the off-mass-shell nucleon requires a different
 description than just a kinematical shift in $x$.
 A systematic
 study  of these off-mass-shell effects started
 recently~\cite{th1,kw}, though some
 effects which do not reduce to the binding effects
 were discussed earlier~\cite{grliuti}.
 Following the tradition of these papers~\cite{th1,kw}
 we call, through out this paper, the effects of the second kind as
 "off-shell" effects, distinguishing them from the binding effects.

 In ref.~\cite{th1} the general form of the
 truncated nucleon tensor (see Fig.~1(a)) is studied.
 The structure functions
 of physical states are then presented as  Feynman
 diagrams with insertion of the truncated nucleon
 tensor.
 For the transverse unpolarized structure functions
 the form of the insertion is found to be:
 \begin{eqnarray}
  m\hat W_T(p,q) = \hat q \chi_1(p,q) + \hat p \chi_2(p,q)+
m\chi_3(p,q),
  \label{tens}
 \end{eqnarray}
 where $\chi_i$ are three different  scalar functions,
 while for the scattering on the point-like fermion only the term
 $\sim \hat q$ exists~\cite{muld}. Calculation of the nuclear
structure
 functions, keeping only the $\sim \hat q$ term leads to the
 existence only of the binding effects and the convolution
 formula~\cite{levin,physlet},
 which is a reasonable approximation, since the
 entire non-relativistic nuclear physics works with point-like
 nucleons. Non-uniform deformation of each of the three terms
 in  eq.~(\ref{tens}) off the mass-shell and
 the non-trivial spin structure of $\hat W_T(p,q)$ lead to
 new effects in the processes involving the off-mass-shell
 nucleons.

\vskip 5mm
\let\picnaturalsize=N
\def\picsize{4.5in}
\def\picfilename{diag1.eps}
\ifx\nopictures Y\else{\ifx\epsfloaded Y\else\input epsf \fi
\let\epsfloaded=Y
\centerline{\ifx\picnaturalsize N\epsfxsize \picsize\fi
\epsfbox{\picfilename}}}\fi
 \begin{center}
\begin{minipage}{15cm}
Figure 1. {\em The truncated nucleon tensor
in the pion-nucleon model. Graph for the
dressed nucleon tensor (a) is the sum of
graphs (b)-(c), where virtual photon
scatters on the bare nucleons and mesons.}
\end{minipage}
\end{center}
\vskip 3mm

 The investigations of refs.~\cite{th1,kw,grliuti} are based on
 models for the nucleon structure, motivated by the constituent
 quark model, and they involve
 unknown
 quark-nucleon amplitudes which are parametrized to fit
 the nucleon structure function on the mass-shell. The
 off-shell effects are found to be only a few percent
 in the absolute value
 of the structure function of nuclei.
 Apparently the magnitude
 and behavior of these corrections are dependent on the
 form of the parametrizations of the quark-nucleon vertices
 with parameters being fixed to describe the
 observables (structure functions) for the nucleon on the mass shell.
 Whereas the continuation to off-mass-shell region must be  ruled
 by dynamics (the equations of motion for participating fields)
 and
 fixed by some constraints, such as
 the Ward-Takahashi relations, etc, so as
 to provide consistency of the model with the calculation
 of any other observables.
 On the other hand, the smallness of the corresponding corrections
 makes it difficult to test models experimentally in the
 usual deep inelastic experiments on nuclei.

 The circumstances mentioned above motivate us to continue
 investigations of the off-shell effects in the nucleon
 structure function. In particular, we aim (i)
 to consider the dynamical model for the nucleon structure
 function other than the quark models of refs.~\cite{th1,kw,grliuti},
 so as
 to compare results of independent approaches to the problem and
 (ii)  to
 discuss possible experiments, other than inclusive DIS,
 which can open new perspectives to study off-shell effects.

  In the present paper we consider the pion-nucleon model
 for the structure function of the nucleon, which is
 motivated by the well-known Sullivan model~\cite{sul,thom,sulpro}.
 We argue that the model is relevant for the case and
 discuss model ambiguities involved in the calculations.
 Then we calculate
 off-shell effects in the inclusive and semi-inclusive
DIS on the deuteron. The second reaction provides new opportunities
 for an experimental investigation of the
 off-shell effects in the nucleon structure function.

{\section {The Pion-Nucleon Model for the Nucleon Structure
Function}}

 The role of the pion cloud or, more generally, mesonic
 cloud in the formation of the
 peripheral structure of the nucleon structure function
 has been  discussed widely~\cite{sul,thom,sulpro,sulpro2}.
 We will discuss only the pion cloud, since its contribution
 significantly dominates over
 the contributions of the heavier mesons.
 The conventional analysis is based on corrections to
 the scattering on the nucleon, calculated
 in the lowest order on the pion-nucleon coupling constant, $\sim
g^2$,
 (see Fig.~1(b)-(d)) with pseudoscalar coupling:
  \begin{eqnarray}
{\cal  L_{\it int}} = - ig\bar \psi (x)
\gamma_5\mbox{$\boldmath{\tau}$}\mbox{$\boldmath{\phi}$}\psi(x),
  \label{li}
     \end {eqnarray}
 Then the diagrams
 (c) and (d) in the pion-nucleon model
 are logarithmically divergent.
 The formal logic of the field theory requires
 renormalization by introducing
 structure functions of the "bare" nucleons and mesons
 and some counter-terms
 in such a way as to provide the correct value of the calculated
structure functions
where the
 A natural normalization point for the counter-terms is the nucleon
mass shell.
 This is enough if we are going to consider off-mass-shell
 behavior of the nucleon structure function.
 However, this is not sufficient
 if we are intending to reach conclusions about contributions of the
mesons
 to the structure function of the free nucleon.

 The physically motivated Sullivan model gives the possibility to
estimate the
contribution
 of the pion cloud  to the free nucleon structure function. In the
modern
 form~\cite{thom,sulpro2} the model is based on the graphs (b)-(d) in
Fig.~1
 and ingredients of the
 model include
 elementary structure functions of bare nucleons and mesons
 and  the meson-nucleon
 vertex formfactors
 in the diagrams (c) and (d).
 The vertex formfactors cut the "unphysical" high momenta of the
 pions and make the contribution of the diagrams (c) and (d) finite.
 Physical picture corresponding to the DIS on the nucleon
 in the Sullivan model is the following.   The nucleon is presented
as a
 superposition of two states, the bare nucleon state
 and  the nucleon plus one pion state.
 As a result, the physical nucleon is the point-like
 bare nucleon ("the nucleon core") surrounded by the extended
 pion cloud. The quark-gluon degrees of freedom in this picture are
 "hidden" in the effective hadron degrees of freedom
 and their presence is displayed through the elementary structure
functions
 of the  bare nucleons and mesons.
 This model is relevant to a study of the nuclear effects in the
 nucleon structure function, since the NN-potential
 can be succesfully defined in the same $g^2$-approximation,
 the one-boson-exchange potential, as the dressing diagrams
 on Fig.~1. At the same time the off-mass-shell behavior
 of the structure functions is governed by the meson-nucleon
 dynamics and, therefore, is consistent with dynamics of the
 deuteron, which we assume to describe also in the
 meson-nucleon model~\cite{bj,bonn,gross,tjond}.

 The vertex formfactor plays the crucial role in the Sullivan model
 of the nucleon structure function~\cite{thom,sulpro2}. Since
 without formfactors the
 one-loop diagrams (c) and (d) are divergent,
 the cut-off parameters control the magnitude of the
 contribution of these diagrams  to the  nucleon structure function.
 On the other hand, for our purpose, to analyse the off-shell
 behavior of the nucleon structure function, we can work
 without formfactors, attributing all divergences to
 the renormalization of the structure functions of the bare
 nucleons and mesons. However, intending to make connection with
other
 calculations, we
 perform all numerical estimates within
 the Sullivan  model
 with formfactors.
 The form of vertex formfactor can be chosen as~\cite{gross}:
 \begin{eqnarray}
 F (p,p') = f((p-p')^2)h(p^2)h(p'^2),
 \label{ff}
 \end{eqnarray}
 where $p$ and $p'$ are  incoming and outgoing nucleon momenta in the
 vertex, respectively, $(p-p')$ is the pion momentum. The formfactors
 are normalized so that:
 \begin{eqnarray}
 f(\mu^2)= 1, \quad h(m^2)= 1,
 \label{ffn}
 \end{eqnarray}
 where $\mu$ is the pion mass. So for the on-mass-shell nucleon
 structure function the  diagram (c) is regularized by single
 formfactor $h(p')$ and diagram (d) by $f((p-p')^2)$.

 We accept the point-like behavior of the bare nucleons, while the
 extended structure is generated by the pion cloud,
 dressing diagrams (c) and (d).
  Calculation of the diagrams (b)-(d) gives (we consider the
structure function
 $F_2$):
  \begin{eqnarray}
&& \hat F_2 (x,p^2) = \int\limits_0^1 dy \left ( \hat f^N(y,p^2)
\tilde{F}_2^N\left (\frac{x}{y}\right)
 + \hat f^\pi(y,p^2) \tilde{F}_2^\pi\left (\frac{x}{y}\right) \right
),
 \label{sf}\\
      && \hat f^{N}(y,p^2)= \frac{\hat q}{2pq}
       \left(\delta (1-y) + f_1^{N}(y,p^2)\right)
      + \hat p  f_2^{N}(y,p^2)
      + m f_m^N(y,p^2) \label{sfn},\\
      && \hat f^{\pi}(y,p^2)= \frac{\hat q}{2pq} f_1^{\pi}(y,p^2)
      + \hat p  f_2^{\pi}(y,p^2)
      + m f_m^\pi(y,p^2),
\label{sfm}
 \end{eqnarray}
where $\hat{q}= q_\mu \gamma^\mu$, etc,
 $\tilde F_2^{N(\pi)}$ is the bare nucleon (pion) structure function.
The term $\delta(1-y)$ arises from the diagram (b), $f_i^N$ from
 (c) and $f_i^\pi$ from (d) ($i =1\ldots 3$).
 The explicit form of the functions
 $f_i^{N,\pi}(y,p^2)$
 is presented in the Appendix A.  If neglect the formfactors,
 these functions satisfy the
 condition
 \begin{eqnarray}
 f_i^N (y,p^2) = f_i^\pi (1-y,p^2),
 \label{sr}
 \end{eqnarray}
 which heuristically can be obtained as a consequence
 of the probabilistic interpretation of the
 structure function~\cite{thom}.
 However, inclusion of the formfactors (different for the diagrams
 (c) and (d)!) breaks the relation (\ref{sr}), which leads to a
 violation of the charge and/or momentum conservation in the
process~\cite{thom}.
The underlying reason is that in the covariant calculations it is
impossible
to introduce "symmetric" formfactor with respect to the nucleon and
meson
momenta.

 Other topic is the choice of the cut-off parameters
 in the formfactors. It is known~\cite{thom,sulpro2} that to have a
reasonable
 physical interpretation of the calculated structure functions
cut-off masses
in the formfactor (\ref{ff}) should
 be significantly smaller than it is found from an analysis of the
meson-exchange potentials~\cite{bj,bonn,gross,tjond}. To regulate the
divergent
diagram (c),
the formfactor is chosen~\cite{thom}:
  \begin{eqnarray}
 f(k^2)= \left (\frac{\Lambda_\pi^2 - \mu^2}{\Lambda_\pi^2 -
k^2}\right )^2,
 \quad \Lambda_\pi = 1\quad GeV, \label{mff}
 \end{eqnarray}
which corresponds to the formfactor for diagram (d):
   \begin{eqnarray}
 h(p^2)= \left (\frac{\Lambda_N^2 - m^2}{\Lambda_N^2 - p^2}\right
)^2,
 \quad \Lambda_N = 1.475\quad GeV, \label{nff}
 \end{eqnarray}
where parameter $\Lambda_N$ is fixed to preserve the baryon charge
(not the
momentum!) conservation. On the other hand, there is no reason to
put
a smaller  cut-off mass in the formfactor corresponding  to the
external nucleon line
 for the diagram (c-d), where cut-off mass should be
$\sim 1.5-2.0$ GeV~\cite{gross}.

 We have to stress once again that for the analysis of the
off-mass-shell
effects
in the nucleon structure functions we do not necessarily  need the
regulating
formfactors
 in the divergent diagrams. We are introducing these formfactor only
to
 relate to the well-known and widely discussed Sullivan model for the
 nucleon structure function.
 The dependence of our results  on the choice
 of the cut-off mass, $\Lambda_{\pi(N)}$, in  the loops in diagrams
(c) and (d)
 is weak compared to the $\Lambda_{\pi(N)}$-dependence of the pion
(nucleon)
  contribution
 to the free nucleon structure function.
  Moreover, to restore  symmetry
between nucleon and pion contributions into
eq.~(\ref{sf})-(\ref{sfm}) we
 use  eq.~(\ref{sr}) to define the nucleon contribution
through the pion one. This way to proceed is supported by
analysis~\cite{thom} using the time-ordered perturbation theory,
where it is shown that eq.~(\ref{sr}) is valid and introducing the
formfactors
in the covariant calculations
damages significantly  only the nucleon contribution.

The free nucleon structure function is defined by inserting  operator
eq.~(\ref{sf}) between Dirac spinors.
The final expression coincides with the
the result of the Sullivan model~\cite{thom,sulpro2}:
  \begin{eqnarray}
 F_2^N (x,p^2=m^2) = N_N\int\limits_0^1 dy \left ( f^N(y,m^2)
\tilde{F}_2^N\left (\frac{x}{y}\right)
 + f^\pi(y,m^2) \tilde{F}_2^\pi\left (\frac{x}{y}\right) \right ),
\label{freen}
 \end{eqnarray}
where $N_N$ is the normalization factor defined by the conservation
of
baryon number:
  \begin{eqnarray}
N_N^{-1}  = 1 + \int\limits_0^1 dy f^N(y,m^2).
\label{freenb}
 \end{eqnarray}
A direct calculation shows that (\ref{freenb}) actually preserves
the baryon number in the one-loop approximation, with
 or without ($\Lambda_\pi \to \infty$) formfactors.
As a basic set of parameters for the calculations we take the
formfactor (\ref{mff})~\cite{thom}. In this case
$N_N^{-1} - 1 \approx 0.24$.
Therefore, for bare structure functions of the nucleons and mesons
we take a fit of the empirical structure functions of
the free nucleons and pions.
This gives a reasonable agreement of the calculated structure
function
(\ref{freen}) with the experimental data,
since admixtures of the corrections (c) and (d)
are not too large.

{\section {The Nucleon Contribution
to the Deuteron Structure Functions}}

Now we are in a position to calculate the nucleon contribution
to the deuteron structure function.
We start with a consideration of the inclusive DIS on the deuteron
(Fig.~2(a)).
The structure function for the deuteron, $F_2^D(x_D)$,
is defined as a matrix element
of the operator eq.~(\ref{sf}).
A consistent way is to use the
covariant amplitude for the deuteron, the Bethe-Salpeter
 amplitude~\cite{tjond,umkh} or relativistic wave
functions~\cite{gross},
with the normalization of the amplitude, based on the deuteron
 charge, in the one-loop
approximation.
This will be done elsewhere. Here to test the method
 we utilize the usual
non-relativistic wave function of the deuteron.
The non-relativistic
reduction of the covariant operator (\ref{sf}) is similar
to earlier approaches~\cite{levin,physlet,kw}.
Here we reduce only the $4 \times 4$,
 Dirac structure
of the operator (\ref{sf}) to the $2 \times 2$, Pauli operators,
keeping all kinematics in the relativistic form.

The deuteron structure function then is defined by:
  \begin{eqnarray}
 &&\!\!\!\!\!\!\!\!\!\!\!\!\!\!\!
F_2^D (x_N) = N_D\int d^4 p \int\limits_0^{M_D/m} d\xi
\delta\left (\xi - \frac{p_+}{m}\right ) \left | \Psi_D(p) \right |^2
\xi \nonumber \\
&&\left \{
\frac{\langle \hat q \rangle}{2pq} \tilde
F_2^{N}\left(\frac{x_N}{\xi}\right)+
\frac{\langle \hat q \rangle}{2pq}
F_2^{(1)}\left(\frac{x_N}{\xi},p^2\right)+
\langle \hat p \rangle F_2^{(2)}\left(\frac{x_N}{\xi},p^2\right)+
\langle m \rangle F_2^{(3)}\left(\frac{x_N}{\xi},p^2\right)
\right \},
\label{freed}
 \end{eqnarray}
where $N_D$ is renormalization constant preserving the conservation
of the baryon number in the deuteron,
$M_D$ is the deuteron mass,
$p_0 = M_D - \sqrt{m^2 + {\bf p}^2}$ is the off-mass-shell
energy of the nucleon, $p_+ = p_0 +p_z$ and
  \begin{eqnarray}
 F_2^{(i)}\left(\frac{x_N}{\xi},p^2\right) = \int\limits_0^{1} dy
\left \{
\tilde F_2^{N}\left(\frac{x_N}{y\xi}\right) f_i^N(y,p^2) +
\tilde F_2^{\pi}\left(\frac{x_N}{y\xi}\right) f_i^\pi(y,p^2)
\right \}, \quad i = 1\ldots 3.
\label{freed1}
 \end{eqnarray}

The brackets $\langle \ldots \rangle$ denotes the non-relativistic
expression for the subsequent operators:
  \begin{eqnarray}
\xi \frac{\langle \hat q \rangle}{2pq} = \left ( 1+\frac{p_z}{m}
\right),\quad
\xi \langle \hat p\rangle = 2p_+ \left ( p_0+\frac{\bf p^2}{m}
\right),\quad
\xi \langle m \rangle = 2m p_+.
\label{nonr}
 \end{eqnarray}

It is anticipated
 that the off-shell effects
are  small in the deuteron. This is a consequence of the fact that,
on the average, the shift
from the mass shell for the nucleon in the deuteron is small:
$ \langle p^2 \rangle \approx m^2 (1  - (3\div 4)\cdot 10^{-2})$.
In heavier nuclei this shift is apparently larger, up to $\sim 10 \%$
for nucleus like iron,
which can lead to a more significant effect.
However, it is interesting to find other possibilities to
investigate the off-shell behavior
of the nucleon structure functions.

\vskip 5mm
\let\picnaturalsize=N
\def\picsize{4.50in}
\def\picfilename{diag2.eps}
\ifx\nopictures Y\else{\ifx\epsfloaded Y\else\input epsf \fi
\let\epsfloaded=Y
\centerline{\ifx\picnaturalsize N\epsfxsize \picsize\fi
\epsfbox{\picfilename}}}\fi
 \begin{center}
\begin{minipage}{15cm}
Figure 2. {\em The deep inelastic scattering on the deuteron:
inclusive {\rm (a)} and semi-inclusive {\rm (b)}.}
\end{minipage}
\end{center}
\vskip 3mm

Let us consider now the semi-inclusive DIS
on the deuteron. The spectator mechanism~\cite{ffss} for this
reaction is
presented in Fig.~2(b). Even for non-relativistic momenta of the
spectator
nucleon the shift from the mass shell of the interacting nucleon can
be
much larger than in heavy nuclei! For instance,
$\sim 20 \%$ for $p_s \sim 300 MeV/c$
and $\sim 40 \%$ for $p_s \sim 500 MeV/c$. Thus semi-inclusive
DIS provides a unique possibility to study the off-shell behavior of
the
nucleon structure function. In particular, the off-shell effects can
be
studied as a function of the shift from the mass shell (or spectator
momentum).

In the non-relativistic approach, with disregard of the off-shell
effects,
the structure function of the
off-shell neutron, $F_2^n(x,p_s)$, measured with the detection of the
proton-spectator
has the form:
  \begin{eqnarray}
 F_2^n(x,p_s) = \left ( 1+\frac{p_z}{m} \right)\left | \Psi (-{\bf
p_s})\right
|^2
 F_2^n\left (\frac{x}{\zeta}\right ),
\label{semcon}
 \end{eqnarray}
where $\zeta = p_+/m$, $p_0 = M_D - \sqrt{m^2 + {\bf p_s}^2}$,
$p_z = -({\bf p}_s)_z$. Dividing data by the flux-factor,
$( 1+p_z/m)$, which is known from kinematics, and plotting
$F_2^n(x,p_s)$ as a function of  $z = x/\zeta$ ($z \in (0,1)$) we
should have
a result proportional to the free neutron structure function,
$F_2^n(z)$, with
coefficient $|\Psi(p_s)|^2$, i.e. at any fixed $p_s$ the ratio of the
measured
structure function to the free neutron structure function should be
constant.
This conclusion has to be changed if there is a non-trivial
$p^2$-dependence of the
nucleon structure function, i.e. if the off-shell effects exist.
The structure function $F_2^n(x,p_s)$ in this case is defined
by the equation similar to eq.~(\ref{freed}) only without integration
over
$p$ and with the correspondent isospin modifications.

 \section{Numerical Results and Discussion}

 Prior to calculating  the deuteron structure functions, let us
qualitatively
discuss the possible phenomena   in the structure functions
of the off-shell nucleon. Omitting  details of the
pseudoscalar coupling of the pions and nucleons we can estimate the
amplitude
of the nucleon to emit the virtual meson as
  \begin{eqnarray}
 {\cal M}(p,p') \propto \frac{1}{k^2 -\mu^2},
\label{est1}
 \end{eqnarray}
where kinematics is defined by
  \begin{eqnarray}
&&p = (p_0,{\bf p}), \quad p' = (E', \frac{\bf p}{2} -{\bf k}), \quad
E = \sqrt{m^2 +   (\frac{\bf p}{2} -{\bf k})^2},
\nonumber \\
&&k = (k_0, \frac{\bf p}{2} +{\bf k}), \quad k_0 = p_0 - E'.
\label{est1k}
 \end{eqnarray}
For simplicity let us compare amplitudes for nucleon with $p_1,\quad
(p_{01}=p_0 < m)$ and
"more virtual" nucleon with $p_2,\quad (p_{02}=(p_0-\Delta), \quad
\Delta >
0)$, and other components of $p_1$, $p_2$ and $p'$ are kept
 the same.  The sign of the combination:
  \begin{eqnarray}
 {\cal M}^{-1}(p_1,p') -{\cal M}^{-1}(p_2,p') \propto -\Delta^2 +
2\Delta(p_o-E') < 0,
\label{est2}
 \end{eqnarray}
controls relative magnitude of these two amplitudes.
Since amplitudes (\ref{est1}) are negative, the relation (\ref{est2})
means
that
absolute value of the amplitude ${\cal M}(p_2,p')$ is larger than
${\cal M}(p_1,p')$. It means that for a nucleon further from the mass
shell
an increase in the emission of virtual pions may be expected.
In accordance with eq.~(\ref{sr}) the role of virtual nucleons
is also increased. On the other hand, eq.~(\ref{freenb})
 implies that weight of the bare component is decreased. The maximums
of both
effective  distributions $f_N(y)$ and $f_\pi(y)$ are at $y < 1$, $y
\sim
0.2-0.3$
for pions and $y \sim 0.7-0.8$ for nucleons, therefore contributions
of both
components are concentrated at smaller $x$ than for the bare
component, where
$f_{bare}(y)\propto \delta(1-y)$. As result,
for the off-shell  structure function we expect an increase
at small $x$ and a decrease at large $x$, compared to the structure
function
with
smaller virtuality. These conclusions depend on the choice of the
pseudoscalar
coupling, the vertex formfactors and the Fermi motion in the
deuteron.

 The parametrizations
of the nucleon structure functions from~\cite{amb} and
pion structure function from~\cite{mes} are used as input.
The Bonn potential wave
function for the deuteron~\cite{bonn} is utilized throughout.
The formfactor for the external nucleon line of the diagrams
(c) and (d) on Fig.~1, is taken of the form~\cite{gross}:
   \begin{eqnarray}
 h(p^2)= \frac{2(\tilde \Lambda_N^2 - m^2)^2}
{2(\tilde \Lambda_N^2 - m^2)(\tilde \Lambda_N^2 - p^2)+(m^2-p^2)^2}.
\label{3ff}
 \end{eqnarray}
Since the role of this formfactor in the diagrams for the reaction
with
external probe
is uncertain, results are presented both with ($\tilde \Lambda_N^2 =
1.65$~GeV)
and without ($\tilde \Lambda_N^2\to \infty$) the formfactor. The
second
case is
our choice for basic set of parameters.

The results for the
deuteron structure function, $F_2^D(x)$, are presented in Fig.~3.
The dotted line presents the calculation  with disregard of
the off-shell effects, the convolution model.
Solid curve is a result
 of calculations with full formulae (\ref{sf})-(\ref{sfm})
with our basic set of parameters. The dashed curve shows the effect
of the
extra
formfactor for the external line of the virtual nucleon. The
additional
formfactor
slightly decreases the off-shell effects, since it "holds" the
nucleon
closer to the mass shell.
These results confirm our estimates presented at the beginning
of the present section. They are also in qualitative agreement with
earlier
results~\cite{kw,grliuti} for $x > 0.3$, where the structure
function of the nucleon in the deuteron suffers additional
suppression
in comparison with the usual convolution model. This mechanism,
indeed, can be
complimentary to the binding effects in explaining the EMC-effect.

Corresponding effective distribution functions of the pions are
presented in
the Fig.~4. "Meanvalue" distribution for the deuteron (dashed curve)
differs
only slightly from the
free nucleon distribution (solid line). At the same time the deuteron
distribution
is very similar to the distribution from the semi-inclusive reaction
at
the spectator
momentum $p_s = 100$ MeV/c, the reason is the mean value of the
nucleon
momentum
in the deuteron is $\sim 100-150$ MeV/c, depending on the potential
model.

\vskip 3mm
\hspace*{-.1cm}
\begin{minipage}{12cm}
\let\picnaturalsize=N
\def\picsize{4.5in}
\def\picfilename{emcrat.eps}
\ifx\nopictures Y\else{\ifx\epsfloaded Y\else\input epsf \fi
\let\epsfloaded=Y
\centerline{\ifx\picnaturalsize N\epsfxsize \picsize\fi
\epsfbox{\picfilename}}}\fi
\end{minipage}
 \begin{center}
\vspace*{-.5cm}
\begin{minipage}{15cm}
Figure 3. {\em The ratio of the deuteron structure function
to the free nucleon structure function.
Curves:
solid - basic set of parameters;
dashed - with additional formfactor for the off-mass-shell
nucleon ($\Lambda_g = 1.65$  GeV);
dotted - convolution model.
}
\end{minipage}
\end{center}
\vskip 3mm

Results for the semi-inclusive reaction with the proton spectator
are shown in Fig.~5. Calculations for the ratio of the
neutron structure function in the semi-inclusive reaction to the
free neutron structure function are presented.
the  This ratio is obtained after exclusion of the
(i) flux-factor $(1+p_z/m)$ and (ii) weight $|\Psi (p)|^2$
from the total structure functions.
If the first one is a procedure well-defined by kinematics of the
reaction, the
second
is rather ambiguous, since the wave function of the deuteron is
strongly model dependent. However, it is worthwhile to compare the
relative
effects
at different $p_s$.
If there is no off-shell effect such a ratio would be
just a constant, $\sim |\Psi (p)|^2$. Otherwise it will have a slope
as shown in  Fig.~5. All curves in
Fig.~5 are scaled  for comparison.
Note also that the flux-factors in the three matrix elements
(\ref{nonr})
 are slightly
different, so after dividing by the factor  $(1+p_z/m)$, the
structure functions remain dependant on
angle, $\theta_s$, of the spectator momentum relative to the
${\bf q}$. This dependence is too
weak to discuss in relation to the possible experiments
and, furthermore, it is not clear
if this angular dependence is just an artifact of the
non-relativistic reduction.
Here we choose $\theta_s = \pi/4$.

\vskip 3mm
\hspace*{-.1cm}
\begin{minipage}{12cm}
\let\picnaturalsize=N
\def\picsize{5in}
\def\picfilename{f_pi.eps}
\ifx\nopictures Y\else{\ifx\epsfloaded Y\else\input epsf \fi
\let\epsfloaded=Y
\centerline{\ifx\picnaturalsize N\epsfxsize \picsize\fi
\epsfbox{\picfilename}}}\fi
\end{minipage}
\vspace*{-3mm}
 \begin{center}
\begin{minipage}{15cm}
Figure 4. {\em The effective distribution of the pion in the nucleon
(the basic set of parameters).
Curves:
solid - free nucleon;
dotted - nucleon in the deuteron;
dashed - nucleon in the semi-inclusive reaction with different
spectator
momentum, $p_s$ (1 - $p_s = 0.1$ GeV/c; 2 - $p_s = 0.3$ GeV/c;
3 - $p_s = 0.5$ GeV/c; 4 - $p_s = 0.9$ GeV/c).
}
\end{minipage}
\end{center}
\vskip 3mm

Dependence of the off-shell effects on the spectator proton momentum
is
shown in Fig.~5(a). There are two competing mechanisms, the increase
of the
structure function at small $x$ and decrease at medium $x$.
To understand such a behavior let us consider the
effective distribution functions of the pions
in the nucleon in the reaction (Fig.~4).
A steady increase of the pion distribution function at small $x$
with an increase of the virtuality of the nucleon is found. At medium
and large
$x$
and high virtualities these distributions have a tendency to vanish.
However,
very
high virtuality, or large spectator momentum $\sim 1$ GeV/c, are
probably
beyond, or
very close to the boundary, the
applicability of the pion-nucleon model and the potential model for
deuteron.

\vskip 3mm
\hspace*{-1.3cm}
\begin{minipage}{16cm}
\let\picnaturalsize=N
\def\picsize{7.50in}
\def\picfilename{semiab.eps}
\ifx\nopictures Y\else{\ifx\epsfloaded Y\else\input epsf \fi
\let\epsfloaded=Y
\centerline{\ifx\picnaturalsize N\epsfxsize \picsize\fi
\epsfbox{\picfilename}}}\fi
\end{minipage}
 \begin{center}
\begin{minipage}{15cm}
Figure 5. {\em The ratio of the neutron structure function,
measured in the semi-inclusive deep inelastic scattering on the
deuteron,
to the free neutron structure function.
{\rm (a)} Dependence on the spectator proton momentum. Curves are
calculated
with
a basic set of parameters:
dotted - $p_s= 0.1$ GeV/c;
solid - $p_s= 0.3$ GeV/c;
dashed - $p_s= 0.5$ GeV/c;
dot-dashed - $p_s= 0.9$ GeV/c. {\rm (b)} Sensitivity to the model
assumptions.
Curves for $p_s= 0.3$ GeV/c: solid - basic set of parameters;
dashed - with additional formfactor for the off-mass-shell
neutron ($\Lambda_g = 1.65$  GeV); dotted - calculation
with the distributions {\rm (\ref{nuc1}-\ref{nuc3})} with nucleon
formfactor inside the loop ($\Lambda_N = 1.475$ GeV).}
\end{minipage}
\end{center}
\vskip 3mm

Fig.~5(b) shows estimates of some of the model ambiguities involved
into
our calculations. In particular,  all calculations give
 qualitatively the same behavior of the off-shell effects.
 However, manipulation with the formfactors may lead to a
suppression of the magnitude of the effect. (Note that calculation
for
dotted curve,  Fig.~4(b), breaks the energy-momentum
conservation in the reaction.)

 \section{Conclusions and Comments}

We have presented model calculations of the off-shell effects in
the deep inelastic scattering on the nucleons. In particular,
\begin{enumerate}
\item Truncated  nucleon tensor has been calculated in the
      pion-nucleon model, motivated by the Sullivan model. The
      formulae explicitly contain the $p^2$-dependence  and allow an
analysis
      of off-shell effects in the nucleon
      structure functions.
\item Nucleon contribution in the deuteron structure function,
$F_2^D$,
      has been calculated, using non-relativistic wave function. The
off-shell
      corrections are found to be rather small, but they can be
complimentary
      to the binding corrections in the explanation of the
EMC-effect.
\item Semi-inclusive deep inelastic scattering on the deuteron has
been
      considered
      in the spectator approximation with the proton spectator in the
final
      state. It is found that this reaction provides new
opportunities to study
      the off-shell effects in the nucleon structure function. Even
at
      non-relativistic momenta of the spectator, the off-shell
effects for the
      struck nucleon are larger than an averaging in heavy nuclei and
order of
      magnitude larger than averaging in the deuteron. This type
      of experiments would help to select models relevant to
      describe the structure functions of nucleons and, therefore,
      the nuclei.
\end{enumerate}

We did not consider here other type of mesonic corrections to the
structure
functions of
deuteron (nuclei), the contributions of meson exchange
currents, which
should be part of
a consistent
analysis of the DIS on  nuclei
as a system of interacting nucleon and meson
fields~\cite{levin,physlet,umkh,thoms}.
However, as soon as the internal degrees of freedom
of the nuclear constituents are ``defrozen", such
analysis becomes a non-trivial problem, since
it is not clear how internal dynamics
of the constituents interferes with
the dynamics of the system.
Anyhow, the physics here can be extremely interesting
and there is much to learn about how to build a composite system from
 composite constituents.

Other type of phenomenon  not considered here and which
can have an affect on the deuteron structure functions at
very small $x$, say $x < 0.1$, is the so-called
nuclear shadowing~\cite{shad,thoms}. These corrections would
cancel (or partially cancel) the enhancement of the deuteron
structure function
(see Fig.~3),
$x \to 0$, generated by the pions~\cite{thoms,unf}.

We would like also to make some comments about the pion
(meson) physics in the DIS on nuclei.
This topic has as a long history  as
studies of nuclear effects in DIS
 starting from the famous EMC-effect~\cite{pions}.
At some
point it was concluded that there are no excess pions  in nuclei.
 It was based on simple estimates of the
pionic contribution to the nuclear structure functions
and probably a more correct conclusion has to be that
something has been overlooked in these calculations.
This point was recently re-examined in an interesting
work~\cite{bub}. Not going into details we would like
to note that physics here can be even more intricate.
For instance, the off-shell effects in the
pion structure function
can be significant as it was found in~\cite{shakin}.

The last comment  is related
to the state of the experiment.
There is an interesting potential
in the study of the DIS
on the deuteron in the semi-inclusive
set up. In particular, the possibility
to study the off-mass-shell behavior of the
nucleon structure function is  really unique.
(Other interesting physics could be studied
as well~\cite{ffss}.)
Such experiments, for instance,
at CEBAF~\cite{kuhn},  would be beneficial
both for the
theory of nuclear effects in the DIS
and, perhaps, for the more fundamental theories
(models) of the structure of the nucleon.


 \section{Acknowledgments}

 Authors thank
 L. Celensa,  F. Gross,
  K. Kazakov, S. Kuhn, S. Kulagin, W. Melnitchouk, C. Shakin,
A. Thomas, W. Van Orden and W. Weise for discussions which clarify
 number of questions.
 The research is supported in part
 by the Natural Sciences and Engineering Research Council of Canada.


 \section{Appendix A}
Light cone variables:
    \begin{eqnarray}
&& p = (p_0, {\bf p}) = (p_+,p_-,{\bf p_\bot}),
\quad p_\pm = p_0 \pm p_z,\nonumber\\
&& p^2 = p_+p_- -{p_\bot}^2, \quad
 pp' = \frac{1}{2}(p_+p'_- + p_-p'_+) - {\bf p_\bot p'_\bot}
   \label{lcv}     \\
 && d^4p = \frac{1}{2} dp_+dp_-d{\bf p_\bot},
 \quad d{\bf p_\bot}= p_\bot d p_\bot  d\alpha
. \nonumber
\end{eqnarray}

The explicit expressions for functions $f_i^{N(\pi)}(y,p^2),\quad i
=1\ldots
3$:

   \begin{eqnarray}
&&\!\!\!\!\!\!\!\!\!\!\!\!\!\!\!      f_1^N(y,p^2) =
-\frac{g^2h^2(p^2)}{16\pi^3}\frac{y}{1-y}\int\limits_0^\infty
       dp'_{\bot}p'_{\bot}\int\limits_0^{2\pi}d\alpha
\frac{h^2(p'^2)}{(p'^2-m^2)^2}
\left \{
2yp^2 + \frac{1}{y}[p'^2-m^2]-2pp'
\right\}
,
    \label{nuc1}\\
&&\!\!\!\!\!\!\!\!\!\!\!\!\!\!\!       f_2^N(y,p^2) =
-\frac{g^2h^2(p^2)}{16\pi^3}\frac{y}{1-y}\int\limits_0^\infty
       dp'_{\bot}p'_{\bot}\int\limits_0^{2\pi}d\alpha
\frac{-yh^2(p'^2)}{(p'^2-m^2)^2},
    \label{nuc2}\\
&&\!\!\!\!\!\!\!\!\!\!\!\! \!\!\!      f_3^N(y,p^2) =
-\frac{g^2h^2(p^2)}{16\pi^3}\frac{y}{1-y}\int\limits_0^\infty
       dp'_{\bot}p'_{\bot}\int\limits_0^{2\pi}d\alpha
\frac{h^2(p'^2)}{(p'^2-m^2)^2},
    \label{nuc3}\\[5mm]
&&\!\!\!\!\!\!\!\!\!\!\!\!\!\!\!       f_1^{\pi}(y,p^2) =
\frac{g^2h^2(p^2)}{16\pi^3}\frac{y}{1-y}\int\limits_0^\infty
       dk_{\bot}k_{\bot}\int\limits_0^{2\pi}d\alpha
\frac{f^2(k^2)}{(k^2-\mu^2)^2}
\left \{
2yp^2 -2pk
\right\}
,
    \label{pi1}\\
&&\!\!\!\!\!\!\!\!\!\!\!\!\!\!\!       f_2^{\pi}(y,p^2) =
\frac{g^2h^2(p^2)}{16\pi^3}\frac{y}{1-y}\int\limits_0^\infty
       dk_{\bot}k_{\bot}\int\limits_0^{2\pi}d\alpha
\frac{(1-y)f^2(k^2)}{(k^2-\mu^2)^2},
    \label{pi2}\\
&&\!\!\!\!\!\!\!\!\!\!\!\!\!\!\!       f_3^{\pi}(y,p^2) =
-\frac{g^2h^2(p^2)}{16\pi^3}\frac{y}{1-y}\int\limits_0^\infty
       dk_{\bot}k_{\bot}\int\limits_0^{2\pi}d\alpha
\frac{f^2(k^2)}{(k^2-\mu^2)^2},
    \label{pi3}
\end{eqnarray}
where $h(p^2)$ and $f(k^2)$ are the model formfactors (see
eq.~(\ref{ff})-(\ref{nff})),
$\pm$-components of $p'$ and $k$ are defined as follows:
   \begin{eqnarray}
&&p'_+ = yp_+, \quad k_+ = yp_+,\nonumber \\[2mm]
&&p'_- = \frac{1}{p_+(1-y)}\left [p^2-\mu^2-p'^2_{\bot}-yp_-p_+ +
2p_\bot p'_\bot \cos{\alpha}\right],
    \label{comp}   \\  [2mm]
&&k_- = \frac{1}{p_+(1-y)}\left [p^2-m^2-k^2_{\bot}-yp_-p_+ +
2p_\bot k_\bot \cos{\alpha}\right],\nonumber
\end{eqnarray}

\newpage

\end{document}